\documentclass[showpacs,prl,twocolumn,floatfix]{revtex4}

\usepackage{graphicx,float,amsmath}

\begin{document}

\title{Spin dependent photoelectron tunnelling from GaAs into magnetic Cobalt}
\author{D. Vu}
\affiliation{Physique de la mati\`ere condens\'ee, Ecole Polytechnique, CNRS, 
91128 Palaiseau, France}
\author{S. Arscott, E. Peytavit}
\affiliation{Institut d'Electronique, de Micro\'electronique et de Nanotechnologie
(IEMN), CNRS UMR8520, Avenue Poincar\'e, Cit\'e Scientifique, 59652
Villeneuve d'Ascq, France}
\author{H.F. Jurca, F. Maroun, P. Allongue, N. Tournerie}
\affiliation{Physique de la mati\`ere condens\'ee, Ecole Polytechnique, CNRS, 
91128 Palaiseau, France}
\author{A.C.H. Rowe}
\email{alistair.rowe@polytechnique.edu}
\affiliation{Physique de la mati\`ere condens\'ee, Ecole Polytechnique, CNRS, 
91128 Palaiseau, France}
\author{D. Paget}
\email{daniel.paget@polytechnique.edu}
\affiliation{Physique de la mati\`ere condens\'ee, Ecole Polytechnique, CNRS, 
91128 Palaiseau, France}

\begin{abstract}
The spin dependence of the photoelectron tunnel current
from free standing GaAs films into out-of-plane magnetized Cobalt films is
demonstrated. The measured spin asymmetry ($A$) resulting
from a change in light helicity, reaches $\pm 6 \%$ around zero applied tunnel bias and drops to $\pm 2 \% $ at a bias of -1.6 V applied to the GaAs. This decrease is a result of the drop in the photoelectron spin polarization that results from a reduction in the GaAs surface recombination velocity. The sign of $A$ changes with that of the Cobalt magnetization direction. In contrast, on a (nonmagnetic) Gold film $A \approx 0\%$. 

\end{abstract}

\maketitle

Since the initial discovery of spin dependent tunnelling between a magnetic
metal and a superconductor \cite{tedrow71} and subsequently between two
magnetic metals, \cite{julliere75, moodera94} spin dependent tunnelling has
been extensively studied in fixed, all-solid junctions. This is because such
studies reveal details of surface magnetism and also because tunnel
junctions, in particular metallic magnetic tunnel junctions, \cite
{nagahama07} are technologically important.\cite{zutic04} Tunnelling from
ferromagnetic and anti-ferromagnetic tips has been successfully employed to
observe magnetic ordering in metals down to the atomic scale.\cite
{wiesendanger09} Similarly, spin polarized tunnelling from ferromagnetic
metals and ferromagnetic semiconductors into nonmagnetic semiconductors has
also been reported in both all-solid junctions \cite{fiederling99} and from
a ferromagnetic tip.\cite{alvarado95} In these cases the transient spin
polarization of the post-tunnel electrons is measured via the circular
polarization of the resulting luminescence. In principle the reverse process
should also be possible. The tunnel current of spin polarized photoelectrons
into a ferromagnetic surface should depend on the relative orientations of
the photoelectron spin to the surface magnetization. This phenomenon was the
basis of Pierce's proposal for GaAs tip spin polarized scanning tunnelling
microscopy (SPSTM).\cite{pierce88} However, despite significant experimental work,
\cite{jansen98,nabhan99} the effect has never been convincingly
demonstrated, with experimental difficulties attributed to parasitic optical
effects yielding apparent spin dependent tunnelling, even on nonmagnetic
surfaces.\cite{jansen99, nabhan99}

\begin{figure}[tbp]
\includegraphics[clip,width=8 cm] {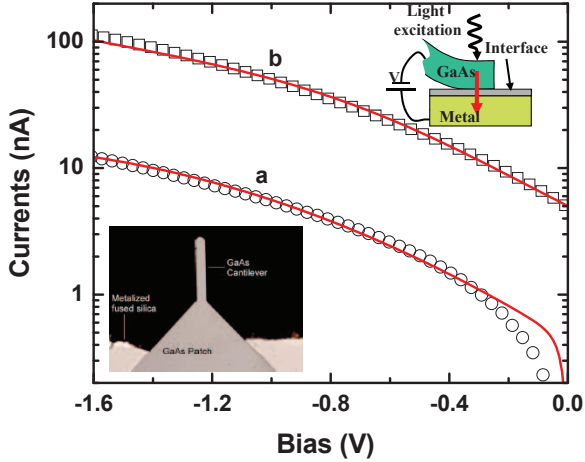}
\caption{The top right schematic describes the principle of the
experiment in which photoelectrons injected from the rear face of a 
free standing GaAs layer diffuse to the front face before tunnelling. 
Also shown, bottom left, is an optical microscope image of the overhanging GaAs layer
deposited onto a metallized quartz substrate. Curves a and b correspond to the tunnel dark and photocurrent bias dependences for
tunnelling into Cobalt, repectively. The solid, red lines correspond to the calculations of the tunnel currents
using a model (Ref. \onlinecite{vu10}) describing tunnelling of photoelectrons into metals.}
\label{fig1}
\end{figure}

Here we demonstrate the spin dependence of the tunnel photocurrent, $
I_{t}^{ph}(\sigma ^{\pm })$, from $p^+$ GaAs under circularly-polarized light
excitation into ultra-thin Cobalt films magnetized out-of-plane. In constrast to previous works\cite{jansen98,nabhan99,jansen99} spin-polarized
electron injection is performed from epitaxial lift-off thin GaAs films
deposited using an original microfluidic method on pre-metallized quartz,
with an overhanging cantilever of $65\mu $m length (see bottom inset, Fig. \ref{fig1}).\cite{arscott10} As shown in the
upper inset of Fig. \ref{fig1}, the photocarriers are generated at the rear (non tunnel) surface, and then diffuse across the film before tunnelling (the film
thickness, of $3\mu $m, is comparable with the charge and spin diffusion
lengths for a doping level of $N_{A}\approx 10^{18}$ cm$^{-3}$ and larger
than the absorption depth, $1\mu $m, for the $h \nu =$ 1.59 eV pump light
used here). The cantilevers
are pressed into mechanical contact with the metal surface, as detected using the reflected part of the incident laser beam with a quadrant photodiode, so that
tunnelling of photoelectrons occurs over a relatively large contact area
through an interfacial oxide layer of homogeneous thickness. This simple,
one-dimensional geometry i) avoids poorly controlled direct light excitation
at the tip apex,\cite{jansen99} ii) results in a photocurrent which,
unlike front surface excitation,\cite{gartner59} does not directly depend on
tunnel bias, iii) reduces instabilities due to changes of interfacial
chemistry observed for tunnelling from tips,\cite{rowe07} and provides a stable tunnel interface for up to 30 minutes in air at room temperature.

\begin{figure}[tbp]
\includegraphics[clip,width=9 cm] {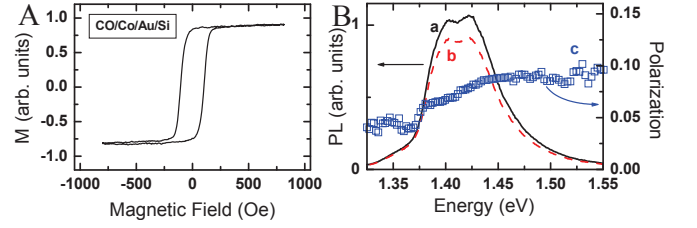}
\caption{(A) The magnetic field dependence of the magnetization
perpendicular to the surface of the Cobalt film as measured using
the polar magneto-optical Kerr effect. 
(B) Curves a and b show the spectra of
the $\sigma^{\pm}$ polarized components of the cantilever
luminescence under circularly-polarized excitation. Curve c shows the
polarization of the spectrum, about $8 \%$ for band-to-band emission.}
\label{fig2}
\end{figure}

Tunnel injection was performed into an ultrathin Co(0001) 
layer (thickness $\approx 5$ monolayers) epitaxially grown by electrodeposition on an atomically flat
Au(111) buffer layer on Si(111).\cite{prodhomme07} The Co surface was
passivated by chemisorbing CO which renders the surface
resistant to oxidation in dry air and quenches empty surface states.\cite{math01} As shown (Fig. \ref{fig2}A) by the square
magnetization loop measured with the field applied perpendicular to the
surface (using the polar magneto optical Kerr effect) averaged over 1 mm$^2$, these passivated Co/Au(111) 
ultra-thin films present a strong perpendicular anisotropy with a coercive field smaller than 200 Oe. The
full zero field remanence of the magnetization after
application of a magnetic field larger than the coercive
field, indicates that the sample is essentially composed of a single domain whose lateral extent is larger
than the contact area through which tunnelling occurs. The photoelectron
polarization in the cantilevers has been analyzed using polarized
luminescence (PL). The $\sigma^{\pm}$ polarized PL spectra $[I_{PL}(\sigma^{\pm})]$ of the cantilever at a low light intensity of
50 W/cm$^{2}$ (h$\nu =$ 1.59 eV) are shown in curves a and b of Fig. \ref{fig2}B, respectively. As known for $p^+$ GaAs, the structure near 1.39 eV
is due to acceptor-related recombination\cite{feng95} and that the above
bandgap luminescence degree of circular polarization, $[I_{PL}(\sigma^+) - I_{PL}(\sigma^-)]/[I_{PL}(\sigma^+) + I_{PL} (\sigma^-)]$ is equal to $8\%$ as
seen from curve c. This polarization corresponds to an average over all
photo-electrons in the cantilever. Using this value
and by numerically solving the spin diffusion equation, a spin polarization
of tunnelling electrons of the order of $16\%$ can be inferred\cite
{arscott10} as well as a spin-lattice relaxation time for conduction
electrons of 0.16 ns, in good agreement with independent measurements on
doped GaAs.\cite{zerrouati88}

\begin{figure}[tbp]
\includegraphics[clip,width=8 cm] {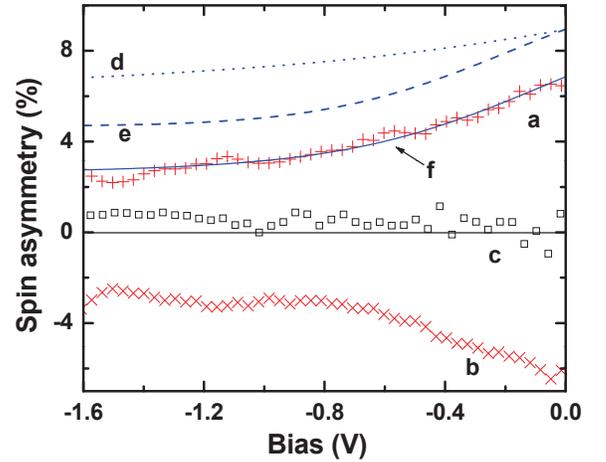}
\caption{Curves a and b show the measured bias dependence of the spin asymmetry of the tunnel
photocurrent into magnetized Cobalt before and after reversal of the magnetization
by the transient application of a magnetic field. Curve c is the asymmetry measured on a nonmagnetic Gold surface.
The calculated spin dependence of the metallic density of states and of the
photoelectron spin polarization, after division by factors of 10 and 1.3
respectively, are shown in curves d and e. The calculated asymmetry, shown
in curve f, is in excellent agreement the measured dependence.}
\label{fig3}

\end{figure}

For the investigation of spin dependent tunnelling, the circular
polarization of the pump light excitation (5 mW focussed to a spot of about $
10\mu$m diameter) is switched by a Pockels' cell. A measurement cycle
consists of the following phases: i) The tunnel current is stabilized at 11
nA in the dark by the feedback loop for a GaAs bias of -1.5 V. ii) The feedback loop is opened and two bias scans of
duration 12 ms are performed. One scan is performed in the dark and the
other one under $\sigma ^{+}$ illumination. The tunnel photocurrent $I_{t}^{ph}(\sigma ^{+})$ is obtained by difference. iii) After a new
stabilization sequence, two bias scans are again taken, one in the dark and
the other one with a $\sigma ^{-}$ polarized laser. This procedure, lasting
about 0.25 s, gives the bias dependence of the spin asymmetry factor $A$,
defined by $A=[I_{t}^{ph}(\sigma ^{+})-I_{t}^{ph}(\sigma
^{-})]/[I_{t}^{ph}(\sigma ^{+})+I_{t}^{ph}(\sigma ^{-})]$. $A$ may also be
written \cite{julliere75} \begin{equation} A=\frac{\delta \rho _{m}}{\rho _{m}}\frac{\delta n_{s}}{n_{s}}, \label{asym}
\end{equation} where $\delta X$ symbolizes the difference of the quantity $X$ between + and
- spins, quantized along the direction of light excitation. $\rho _{m}$ and $
n_{s}$ are respectively the total metallic density of states at the tunnel energy
and the concentration of the tunnelling electrons. Using, as shown above, $\delta n_s/n_s \approx 16\%$, and $\delta \rho_m/\rho_m \approx 70 \%$ about 1 eV above the Fermi energy\cite{math01}, an asymmetry of
the order of 10 $\%$ is anticipated using Eq. \ref{asym}.

\begin{figure}[tbp]
\includegraphics[clip,width=9 cm] {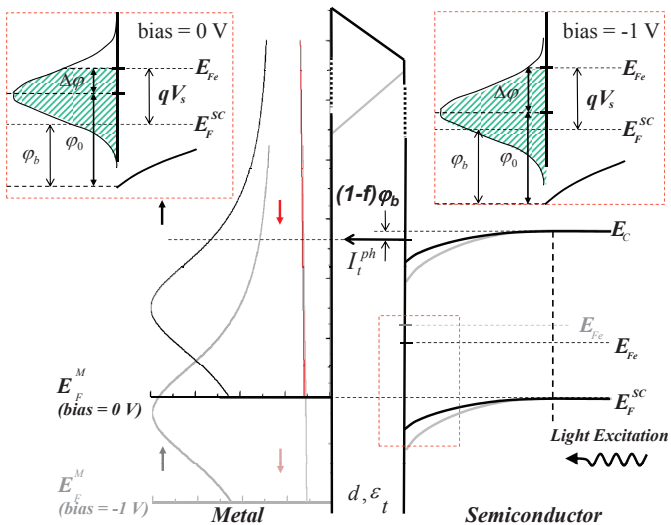}
\caption{Energy band structure for spin polarized tunnelling into Cobalt. The injection
energy $E_g - (1-f)\protect\varphi_b$ is shown along with realistic
representations of the densities of states of the majority (black) and
minority (red) spins. The black (gray) lines represent the case where the applied bias is
0V (-1 V) and indicates that the spin-polarized conduction electrons are injected into the
top half of the 3d minority spin band. The top left (right) inset shows the semiconductor surface
density of states at an applied bias of 0V (-1 V).}
\label{fig4}
\end{figure}

The results averaged over 100 measurement cycles are shown in Fig. \ref{fig1}
and Fig. \ref{fig3}. Curve a of Fig. \ref{fig1} shows the dependence of the
dark current as a function of reverse bias applied to the GaAs cantilever
and curve b shows that of the additional current, $I_{t}^{ph}$, induced by
the light excitation. This current increases nonexponentially up to about 100
nA. Curve a of Fig. \ref{fig3} shows that $A$ varies from $6 \%$ at zero bias
to $2 \%$ at a reverse bias of -1.6 V. The non zero value of $A$ is due to a
spin dependence of the tunnelling current since i) reversal of the
magnetization of the Cobalt layer by transient application of a magnetic
field larger than the corecive field induces a change of sign of the asymmetry without any significant
modification of either the absolute value or the bias dependence (curve b in
Fig. \ref{fig2}), and ii) measurements on (nonmagnetic) Gold films result in
an asymmetry that is always smaller than $1 \%$ (curve c) and approximately $
0 \%$ for zero bias. Moreover the measured asymmetry is similar to the above rough estimate.

A more quantitative interpretation of these results uses a general model recently developed for
tunnel injection of photoelectrons into metals.\cite{vu10} The excellent agreement between the calculated (red lines, Fig. \ref{fig3}) and measured bias dependences indicates that the dominant contribution to the tunnel
photocurrent comes from conduction electrons. The injection energy is almost bias-independent and close to that of the bottom of the
conduction band in the bulk since the energy loss, $(1-f)\varphi _{b}$, in the depletion layer (see Fig. \ref{fig4}) is smaller than 150 meV. (The surface barrier $\varphi _{b}\approx 0.3eV$ under light excitation and the 
numerical factor $f$ is larger than about 0.5 because of surface quantization.)
The energy dependence of the total Cobalt density of
empty states at this injection energy cannot explain the nonexponential bias dependence of
$I_{t}^{ph} $.\cite{math01} In the same way, as shown in curve d of Fig. \ref
{fig3}, $\delta \rho _{m}/\rho _{m}$ calculated using the known spin
dependent density of empty states, only decreases by $25\%$ which, using Eq. 
\ref{asym}, cannot explain the measured bias dependence of $A$. The decrease
of $A$ must therefore be dominated by $\delta n_{s}/n_{s}$.

The decrease of $\delta n_{s}/n_{s}$ and the nonexponential increase of the
tunnel photocurrent are caused by the same effect, namely unpinning of the
surface Fermi level.\cite{vu10} As seen in Fig. \ref{fig4}, the application
of a bias changes the semiconductor surface charge and shifts the electron
quasi-Fermi level away from midgap by a quantity $\Delta \varphi$ which is obtained by charge neutrality.\cite{vu10} The surface
recombination velocity is $S=S_{0}\exp (-\Delta \varphi
/k_{B}T)/D(\Delta \varphi )$ where $S_{0}$ is the value of $S$ for $\Delta \varphi 
$=0 and $D(\Delta \varphi )$ is the relative decrease of the density of surface
states.\cite{comment1prl10} The resulting bias-induced decrease of $S$ \
results in an increase of the effective lifetime of the tunnelling
electrons which increases their concentration and increases the spin polarization losses by spin-lattice relaxation.

The tunnel photocurrent is proportional to $n_{s}$ and to the tunnel probability, 
for which the expressions are found in Ref. \onlinecite{vu10}. Calculation of $\delta
n_{s}/n_{s}$ is performed by solving the equations for spin and charge
diffusion\cite{favorskiy10} from the rear surface to the plane of injection.
For a cantilever of thickness $l$, in the limit of large recombination at the
rear surface and of small absorption length, one finds 
\begin{equation}
\frac{\delta n_{s}}{n_{s}}=\pm 0.5\sqrt{\frac{\tau _{s}}{\tau }}\frac{\sinh
(l/L)}{\sinh (l/L_{s})}\frac{1+S/v_{d}}{1+aS/v_{d}}  \label{deltans}
\end{equation}
for $\sigma ^{\mp }$ polarized light excitation, respectively. Here $\tau$
and $\tau _{s}$ are the bulk electron lifetime and spin lifetime, $L$ and $
L_{s}$ are the charge and spin diffusion lengths, $v_{d}=(D/L)\coth (l/L)$ is
the effective charge diffusion velocity, $D$ is the diffusion constant and $
a=(L_{s}/L)\coth (l/L)/\coth (l/L_{s})$ is the ratio of $v_{d}$ to the
equivalent spin diffusion velocity (here $a<1$ since $L_{s}<L$).

The bias dependences of the tunnel photocurrent, of the dark current and of $A$
are calculated using the model of Ref. \onlinecite{vu10}. The work function
for passivated Cobalt is 6 eV,\cite{ishi85} and the dielectric constant of
the tunnel gap is equal to 10, close to that of both Gallium Oxide\cite
{passlack94} and Cobalt Oxide.\cite{rao65} The spin diffusion length is $
0.6\mu $m, i.e. close to independent estimates.\cite{vu10b} As in Ref.
\onlinecite{vu10}, other parameters for non polarized tunnelling have values taken
from the literature. Good agreement with the data is obtained for $0.6 < f < 1$
when the tunnel distance is adjusted between
\ 0.6 nm and 0.75 nm. The calculated curves in Fig. \ref{fig1} and Fig. \ref{fig3}
correspond to $f \approx 0.9$ and $d = 0.74$ nm. As
seen in Fig. \ref{fig1}, the calculation correctly predicts the bias
dependence of the tunnel dark current and photocurrent. Note that these dependences appear to be quite similar since both are determined by the degree of unpinning of the semiconductor surface Fermi level. Curve e of Fig.\ref
{fig3} shows the calculated decrease of the polarization of injected
electrons. The bias dependence of $A$ calculated using Eq.
\ref{deltans} is shown in curve f and agrees very well with the measured dependence. The zero bias asymmetry is also well accounted for, and is smaller than the rough estimate made above because $\Delta \varphi$ is non negligible for the high excitation intensities used here. 

We have neglected here the spin dependence of the photovoltage and therefore
of $\Delta \varphi$,\cite{jansen98} caused by spin injection into the
subsurface depletion layer. This should induce a spin dependence of the
surface recombination velocity which, as for bulk spin dependent
recombination,\cite{paget84} increases $\delta n_s/n_s$. Conservation of
spin currents shows that the relative change of $\delta n_s/n_s$ depends on
the balance between the spin lattice relaxation (time $T_{1s}$) and the
lifetime of electrons trapped at surface centers. An upper limit for this
effect, found by taking for $T_{1s}$ equal to the spin relaxation time of
conduction electrons (0.16 ns), and a hole capture cross section $\sigma_p =
2 \times 10^{-18}$ m$^2$ equal to the maximum value obtained for a large
variety of midgap centers,\cite{henry77} indicates that the relative modification of the spin asymmetry is less than $10^{-3}$.
Finally, a possible spin dependence of the tunnel matrix element has also
been neglected. While such a dependence is unknown, the good agreement
between the model and the experimental results of Fig. \ref{fig2} indicates that
it does not play a crucial role.

In conclusion, the spin dependence of the tunnel current of conduction
photoelectrons into a magnetic metal has been clearly demonstrated. In
mechanical contact, the bias dependence of $A$ is
caused by the decrease of the electron spin polarization due to the decrease
of the surface recombination velocity resulting from the unpinning of the
quasi electron Fermi level. Spin injection concerns electrons of
well-defined energy (comparable with $k_BT$) and this observation may
finally, for larger tunnel distances where the surface recombination
velocity is nearly constant,\cite{vu10} open the way to spin-dependent
tunnelling spectroscopy (SPSTS) and SPSTM of magnetic metals as
proposed by Pierce more than 20 years ago.\cite{pierce88}

\section{Acknowledgements}

This work was partially supported by the Agence National de la Recherche
(ANR), SPINJECT 06-BLAN-0253.

\bibliographystyle{apsrev}
\bibliography{bibrowe}

\end{document}